\documentclass[journal,10pt]{IEEEtran}
\usepackage{graphicx, epstopdf}
\usepackage{amssymb, amsmath}
\usepackage{subeqnarray}
\usepackage{cases}
\usepackage{commath}
\usepackage{setspace}
\usepackage{acronym}
\usepackage{mathrsfs}
\usepackage{ifthen}
\usepackage{textcomp}
\usepackage{float}
\usepackage{subfigure}
\usepackage[table]{xcolor}
\usepackage{color,soul}
\usepackage{cite}
\usepackage{setspace}
\usepackage{mathtools}
\usepackage{bm}
\usepackage{url}
\usepackage{xcolor}
\usepackage{empheq}
\usepackage{tikz}
\usepackage{subfigure}
\usepackage{stfloats,amsthm}
\usepackage{algorithm}
\usepackage{algpseudocode}
\newtheorem{theorem}{Theorem}

\DeclareMathSizes{12}{12}{7.7}{5.5}
\usepackage[bb=boondox,bbscaled=1]{mathalfa}

\def\argmin{\mathop{\rm argmin}}

\begin{document}

\title{Fast Randomized-MUSIC \\ for mm-Wave Massive MIMO Radars}

\author{Bin~Li$^{1,2}$,
        Shuseng~Wang$^{3}$,
        Jun~Zhang$^{2,4}$,
        Xianbin~Cao$^{4}$,
        Chenglin~Zhao$^{1}$

\thanks{\emph{1} School of Information and Communication Engineering,
Beijing University of Posts and Telecommunications, Beijing, 100876, China.}
\thanks{\emph{2} School of Information and Electronics, Beijing Institute of Technology, Beijing, 100081, China.}
\thanks{\emph{3} Department of Computer Science, Stevens Institute of Technology, Hoboken, NJ 07094, USA.}
\thanks{\emph{4} School of Electronic and Information Engineering, Beihang University, Beijing, 100191, China.}
}

\maketitle

\begin{abstract}
Subspace methods are essential to high-resolution environment sensing in the emerging unmanned systems, if further combined with the millimeter-wave (mm-Wave) massive multi-input multi-output (MIMO) technique.
The estimation of signal/noise subspace, as one critical step, is yet computationally complex and presents a particular challenge when developing high-resolution yet low-complexity automotive radars.
In this work, we develop a fast randomized-MUSIC (R-MUSIC) algorithm, which exploits the random matrix sketching to estimate the signal subspace via approximated computation.
Our new approach substantially reduces the time complexity in acquiring a high-quality signal subspace.
Moreover, the accuracy of R-MUSIC suffers no degradation unlike others low-complexity counterparts, i.e. the high-resolution angle of arrival (AoA) estimation is attained.
Numerical simulations are provided to validate the performance of our R-MUSIC method.
As shown, it resolves the long-standing contradiction in complexity and accuracy of MIMO radar signal processing, which hence have great potentials in real-time super-resolution automotive sensing.
\end{abstract}


\IEEEpeerreviewmaketitle

\section{Introduction}
For the emerging unmanned aircrafts or vehicles, the high resolution environment sensing is critical to safety and operational efficiency~\cite{Alonzo2004Toward}, e.g. collision avoidance, target tracking and etc.
Unlike other automotive radars, e.g. lidar, ultrasound or camera radars, a millimeter-wave (mm-Wave) radar is immune to various adverse conditions (e.g. dust, fog or light) and is regarded as one promising candidate~\cite{waldschmidt2014future}.
Given the recent advances in mm-Wave hardware \cite{hasch2012millimeter}, virtual massive multiple input multiple output (MIMO), i.e. co-located MIMO \cite{li2007mimo}, can be readily deployed to enable high-resolution automotive sensing.
As shown \cite{bilik2018automotive}, for one co-located MIMO radar with $P$ transmitting and $Q$ receiving elements, the equivalent receiving channels would be $M=P\times Q$.
By further cascading multiple chips, e.g. recently developed 4-chip cascaded automotive radars \cite{bilik2018automotive}, the number of equivalent receiving antennas may approach several hundred (e.g. $M=$196), which greatly enhances the spatial resolution of automotive sensing.

As far as the high resolution target estimation is concerned, the subspace methods, represented by MUSIC and ESPRIT \cite{ schmidt1986multiple, roy1989esprit}, have inspired the widespread interest.
By identifying signal/noise subspace via eigenvalue decomposition (EVD) or singular value decomposition (SVD), the near-optimal target estimation can be attained.
Despite the greatly improved resolution, massive MIMO radar (e.g. $>$200 channels) requires a high complexity to acquire signal/noise subspace.
For real-time automotive sensing via massive MIMO radars, such super-resolution methods tend to be impractical, as they incur a huge computation and an intolerable delay.
%

Various algorithms have been designed to alleviate the high complexity in subspace estimation.
By introducing a Propagator operator \cite{marcos1995the}, the computational SVD is replaced by a simple least-squares (LS) regression.
Despite its lower complexity, this Propagator method produces low-resolution or even uncertain results, especially for the angular range ([70, 90] or [-90 -70] degree) \cite{tayem2005l-shape}.
Recently, a similar LS concept is extended to 3-D tensor cases, and 2-D direction of arrival (DoA) information is acquired via one parallel factor (PARAFAC) method \cite{wen2020joint}.
Although the spatial resolution was improved, its computational complexity is rarely feasible to massive MIMO radars.
In ref. \cite{oh2015low}, the inverse of covariance matrix is used to approximate subspace, which, however, has still an unaffordable complexity and produces unstable results when signal-to-noise ratio (SNR) is low.
Other simple methods, e.g. the FFT method, are available, which is yet inadequate to the high-resolution automotive sensing~\cite{patole2017automotive}.

In this paper, we present a new randomized-MUSIC (R-MUSIC) method, which enables low-complexity and high-resolution target estimation in the emerging automotive sensing.
To resolve a long-standing contradiction between complexity and accuracy, we exploit a randomized low-rank approximation (LRA) method to approximately estimate the signal subspace.
Then, the time complexity is dramatically reduced, which is much faster than an exact computation of SVD of large covariance matrix (as in classical MUSIC).
As shown, our new method significantly accelerates the DoA estimation, which causes no accuracy degradation and acquires the near-optimal resolution as classical MUSIC.
Our R-MUSIC method provides the great potential to the real-time and high-resolution automotive sensing with massive MIMO radars.

\section{System Model}
Owing to the low cost radio-frequency (RF) processing \cite{bilik2018automotive,patole2017automotive,oh2015low}, massive MIMO radars, combined with the frequency modulated continuous wave (FMCW) technique, are regarded as one promising solution to high-resolution environment sensing in autonomous systems.
In this section, we briefly introduce the signal model, and discuss some popular algorithms for unknown DoA estimation.

\subsection{FMCW Massive MIMO Radar}

Without loss of generality, we consider a uniform linear array (ULA) massive MIMO radar, which has $M$ receiving elements.
In practice, both real MIMO and virtual MIMO can be deployed in the front bumper of automotive vehicles.
For the $m$th element ($m=0,1,\cdots,M-1$), the received signal of in the line-of-sight (LoS) case\footnote{In this work, we focus on the fast subspace computation and the high-resolution DoA estimation. For the more complicated scenarios, one may refer to other works dealing with related practical challenges, e.g. NLoS propagation or correlated signals \cite{ariananda2012direction,cheng2019doa}, whereby our proposed method would be directly applied to accomplish the fast subspace estimation.} is:

\begin{small}
\begin{align}
y_m(t) =  \sum_{k=0}^{K-1}\alpha_{k,m} s (t-\tau_k) \cdot \text{exp} \left[ -j 2\pi/\lambda_s m d \text{sin}(\theta_k) \right] + n_m(t), \nonumber
\end{align}
\end{small}%
whereby the emitted FMCW signal reads:
\begin{align}
s(t)= \text{exp} \left[ j\left(w_s t +  \mu t^2/2 \right) \right], ~~~0\leq t < T_{\text{sym}}.
\end{align}
In the above, $w_s$ is the initial frequency, and $\mu$ is the changing rate of the instantaneous frequency. Given the chirp bandwidth $w_B$ and a repetition interval $T_{\text{sym}}$, the changing rate is $\mu=w_B/T_{\text{sym}}$.
Here, $\tau_k$ and $\theta_k$ are time-of-arrival (ToA) and DoA of the $k$th target ($k=0,1,\cdots,K-1$), respectively.
$\alpha_{k,m}=\alpha_{k}$ is the complex channel gain, as in common far-field scenarios.
$\lambda_s$ is the wavelength, and $d$ is the spacing distance of two elements (in practice we usually have $d=\lambda_s/2$). $n_m(t) \thicksim \mathcal{N}(0,\sigma_{n,m}^2)$ denotes the additive white Gaussian noise (AWGN), with a variance $\sigma_{n,m}^2=\sigma_n^2$.

Note that, deploying hundreds of antennas would incur the formidable hardware complexity. Thus, massive MIMO can be implemented via a co-located MIMO technique in practice.
Assume we have $M_R$ receiving antennas and $M_T$ transmitting antennas.
Taking a general time division multiplexing (TDM) mode for example, each transmitter antenna would emit the signal at one time slot.
E.g., the 1st antenna may transmit signal at the 1st time slot, whilst all $M_R$ antennas receives the signal.
After total $M_T$ non-overlapped slots, the received signals can be reformatted to a $M_T M_R\times 1$ vector.
Provided the phase information has been calibrated, the equivalent number of receiving elements then becomes $M=M_T M_R$.
Furthermore, if multiple sub-chips are cascaded, the massive MIMO receiver can be easily implemented.

\subsection{Targets Detection and Estimation}

For massive-MIMO radars, usually the high-resolution target estimation can be realized via the spatial pseudo-spectrum \cite{waldschmidt2014future,patole2017automotive} computed from the signal/noise subspace, with which multiple DoAs are then estimated.
Similarly, the ToAs (related to ranges) are attained via the temporal pseudo-spectrum \cite{oh2015low}.
In both cases, the signal (or noise) subspace needs to be exactly estimated, e.g. via either EVD/SVD \cite{schmidt1986multiple,roy1989esprit} or other alternatives \cite{marcos1995the,oh2015low}.

In this analysis, we focus on the estimation of the spatial DoA information, and develop a high-resolution yet low-complexity signal-subspace estimation scheme.
To begin with, a spatial covariance matrix (SCM), which is one symmetric positive semi-definite (SPSD) matrix, is obtained by:
\begin{align}
\textbf{R} =  \mathbb{E}\{\textbf{Y} \textbf{Y}^H\} \simeq
\frac{1}{N} \textbf{Y} \textbf{Y}^H,
\end{align}
where $\textbf{Y} \in \mathbb{C}^{M \times N}$ denotes a signal matrix after de-chirping; $N$ is the sampling length.
While the methods requiring small $N$ or even a single snapshot (i.e. $N=1$) exist \cite{liao2016music}, they may produce less accurate estimation.
Here, we assume $N\thicksim \mathcal{O}(M)$ for the purpose of the high-resolution sensing.
In the following, we also consider $\textbf{R}$ is low ranked, i.e. $\text{rank}\{\textbf{R}\} = K \ll M$, as in most practical applications.

\subsubsection{{SVD based Subspace}}
As for the near-optimal MUSIC scheme \cite{schmidt1986multiple,patole2017automotive}, EVD or SVD is firstly applied to the above SCM $\textbf{R}$. On this basis, two subspaces (i.e. signal and noise) will be identified, i.e.,
\begin{align}
\textbf{R} &= \textbf{U} \pmb{\Sigma} \textbf{V}^H = \textbf{U} \begin{bmatrix}
\pmb{\Sigma}_s & \textbf{0}_{K \times (M-K)}\\
\textbf{0}_{ (M-K)\times K} & \sigma_n^2\textbf{I}_{M-K}
\end{bmatrix}  \textbf{V}^H  \nonumber \\
&= \textbf{U}_s \pmb{\Sigma}_s \textbf{U}_s^H + \sigma_n^2 \textbf{U}_e \textbf{U}_e^H,
\end{align}
where $\textbf{U}\in \mathbb{C}^{M \times M}$ and $\textbf{V} \in \mathbb{C}^{M \times M}$ are respectively the left and right singular matrix; and for a Hermitian matrix $\textbf{R}^H=\textbf{R}$, we may have $\textbf{U}=\textbf{V}$;
$\sigma_R (0) \geq \sigma_R (1)  \cdots \geq \sigma_R(M-1) \geq 0$ are the singular values;
$\pmb{\Sigma}=\text{diag}\{[\sigma_R(0)~\sigma_R(1)~\cdots~\sigma_R(M-1)]^T\} \in \mathbb{R}^{M \times M}$ is a diagonal matrix.
Thus, the signal subspace $\textbf{U}_s$ and the noise subspace $\textbf{U}_e$ can be estimated,
\begin{align}
\textbf{U}_s &\triangleq  [\textbf{u}_0~\textbf{u}_1~\cdots~\textbf{u}_{K-1}] \in \mathbb{C}^{M \times K}, \\
\textbf{U}_e &\triangleq  [\textbf{u}_{K}~\textbf{u}_{K+1}~\cdots~\textbf{u}_{M-1}] \in \mathbb{C}^{M \times (M-K)}.
\end{align}
Finally, for a set of grid angle $\pmb{\theta}=[\theta_0~\theta_1,\cdots,\theta_{L-1}] \in [0,\pi]^L$, the spatial pseudo-spectrum is calculated via:
\begin{align}
P_{\text{music}}(\theta_l) = \frac{1}{\textbf{a}(\theta_l)\textbf{U}_e^H \textbf{U}_e \textbf{a}^H(\theta_l)},
\end{align}
where $$\textbf{a}(\theta)\triangleq [1~\text{exp}(j2\pi d \text{sin}(\theta)/\lambda)~\cdots~\text{exp}(j2\pi (M-1) d \text{sin}(\theta)/\lambda))],$$ is the steering vector.
If there involve $K$ targets, then the pseudo-spectrum $P_{\text{music}}(\theta)$ should contain $K$ peaks which corresponds to the target DoAs.
By identifying $K$ peaks, unknown DoAs are then extracted.
Note that, the other ESPRIT method also adopts SVD to identify a signal subspace~\cite{roy1989esprit}.
The time complexity of the above procedure is $\mathcal{O}(M^3)$.

\subsubsection{{Other Sub-space Methods}}
Other methods of subspace estimation include the matrix inverse method \cite{oh2015low} and the Propagator algorithm \cite{marcos1995the}.
The matrix inversion algorithm is only applicable in high-SNR scenarios, whilst it has the same complexity as MUSIC.
The Propogator algorithm computes the approximated subspace via a time complexity $\mathcal{O}(MNK)$;
however, it unfortunately results in low-resolution and uncertain results in certain ranges of $\theta$.
Besides, the block-Lanczos algorithm can be also applied to compute exact $K$-SVD, e.g. with the lower complexity $\mathcal{O}(K M^2)$ \cite{Grimes1994A}.
To sum up, the contradicting requirement on low-complexity and high-accuracy presents a particular challenge, which remains one long-standing difficulty for MIMO radar signal processing.

\section{Randomized MUSIC with LRA SVD}

\subsection{Low-Rank Approximation}

Instead of computing the exact SVD, here we resort to the approximated $K$-SVD method, which is inspired by a novel concept of random matrix sketching \cite{clarkson2013low,Li2020Randomized}.
To achieve this, we first obtain the low-dimensional abstract of a covariance matrix $\textbf{R}\in \mathbb{C}^{M \times M}$, i.e.
\begin{align}
\textbf{C} = \textbf{R} \textbf{S},
\end{align}
where the low-dimension matrix $\textbf{C} \in \mathbb{C}^{M \times s}~(K<s\ll M)$ is referred as one sketch, which preserves much useful information in $\textbf{R}$.
$s$ is one user-specific parameter to trade accuracy and complexity.
The sketching matrix $\textbf{S}\in \mathbb{R}^{M \times s}~(s\ll M)$, used to abstract the low-dimension representation $\textbf{C}$, is one random Gaussian matrix.
I.e. each element of $\textbf{S}$ is i.i.d.\ drawn from one Gaussian distribution, $S(i,j)\thicksim \frac{1}{\sqrt{s}}\mathcal{N}(0,1)$.
\begin{figure}[!t]
\centering
\includegraphics[width=8.5cm]{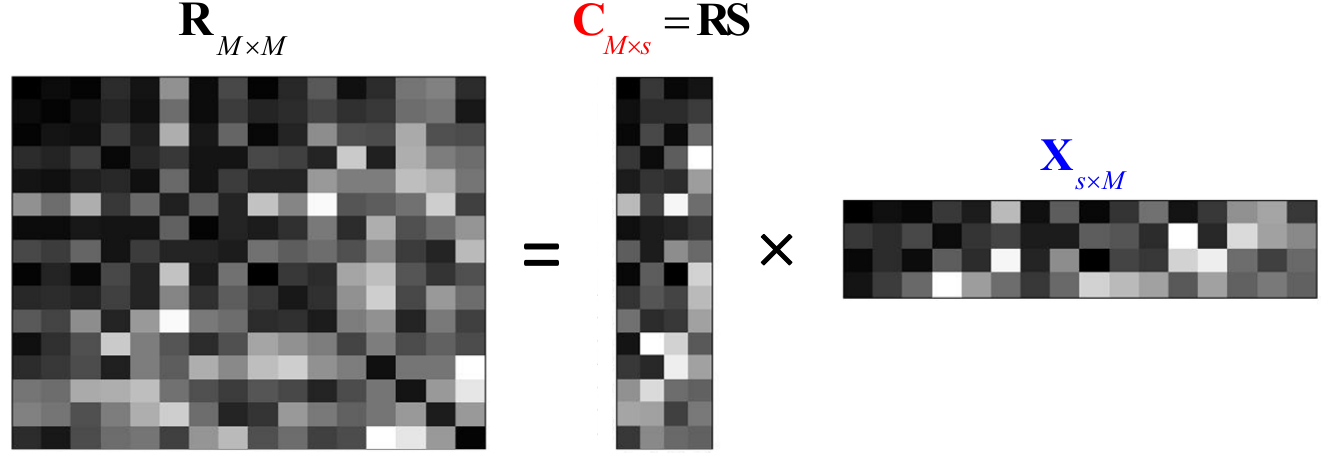}
\caption{Random low-rank approximation inspired matrix factorization.}
\end{figure}

On this basis, we are able to further obtain the low-rank decomposition of $\textbf{R}$, namely $\textbf{R } \simeq \textbf{C} \textbf{X}$. Accordingly, the optimal factorized matrix $\textbf{X}$ is determined by:
\begin{align}
\textbf{X} =  \argmin _{\text{rank}(\textbf{X})\leqslant K}||\textbf{C} \textbf{X} - \textbf{R}||_F^2.
\end{align}
Or alternatively, we can derive the factorized matrix via \cite{Halko2010Finding}:
\begin{align}
\textbf{X} = \argmin _{\text{rank}(\textbf{X})\leqslant K}||\textbf{Q}_{\textbf{C}} \textbf{X} - \textbf{R}||_F^2 =  \textbf{Q}_{\textbf{C}}^T \textbf{R},
\end{align}
where $\textbf{Q}_{\textbf{C}}$ is an orthonormal bases of $\textbf{C}$. 

Based on the above LRA of large covariance matrix $\textbf{R}$, we are able to further obtain its $K$-rank SVD in an approximated (rather than exact) manner:
\begin{align}
\textbf{R}_K &\simeq  \textbf{Q}_{\textbf{C}} \textbf{X} = \textbf{Q}_{\textbf{C}} (\textbf{Q}_{\textbf{C}}^T \textbf{R})_K, \\
& = \textbf{Q}_{\textbf{C}} (\textbf{U}_X \pmb{\Sigma}_X \textbf{V}_X^H) = (\textbf{Q}_{\textbf{C}}\textbf{U}_X) \pmb{\Sigma}_X \textbf{V}_X^H =  \textbf{U}_R \pmb{\Sigma}_R \textbf{V}_R^H,\nonumber
\end{align}
where $\textbf{R}_K$ denotes the best rank-$K$ approximation of $\textbf{R}$, which approximates $\textbf{R}$ via its first $K$ singular values/vectors; and similarly, $(\textbf{Q}_{\textbf{C}}^T \textbf{R})_K$ gives the best rank-$K$ approximation of $(\textbf{Q}_{\textbf{C}}^T \textbf{R})$.
From eq. (10), we have $\textbf{U}_R \triangleq \textbf{Q}_{\textbf{C}}\textbf{U}_X \in \mathbb{C}^{M \times K}$, $\pmb{\Sigma}_R = \pmb{\Sigma}_X \in \mathbb{R}^{K \times K}$ and $\textbf{V}_R=\textbf{V}_X \in \mathbb{C}^{M \times K}$.

\subsection{Approximated Low-Rank SVD}
Note from Eq. (10), the random matrix sketching inspired LRA, namely $\textbf{R}\simeq \textbf{C} \textbf{X} $, potentially allows us to obtain the approximated SVD with a greatly reduced complexity.
Even so, the complexity in calculating the optimal factorized matrix $\textbf{X}$ would be still much high.
Rather than solving the original over-determined system in Eq. (8), we alternatively address it by sketching both a covariance matrix $\textbf{R}$ and its LRA (i.e. $\textbf{C}\textbf{X}$).
As a result, we tend to solve the following problem,
\begin{align}
\hat{\textbf{X}} = \argmin _{ \text{rank}(\textbf{X})\leqslant K}||\textbf{S}_X^T (\textbf{C} \textbf{X} -  \textbf{R})||_F^2,
\end{align}
where the $M \times s_1$ sketching matrix $\textbf{S}_X$ is composed of one Gaussian projection matrix and another Count sketch matrix,
\begin{align}
\textbf{S}_X = \textbf{S}_C \textbf{S}_G \in \mathbb{R}^{M \times s_1}.
\end{align}
Here, $\textbf{S}_C \in \mathbb{R}^{M\times s_0}$ and $\textbf{S}_G \in \mathbb{R}^{s_0 \times s_1}$ denote the count and Gaussian sketching matrices, respectively;
$s_0$ and $s_1$ are two sketching lengths ($K\leq s<s_1<s_0$).
As one popular tool in streaming data computation, the count sketch can be efficiently applied to any matrix.
In principle, a count sketching matrix $\textbf{S}_C$ has two features: (1) each row has only one non-zero entry and its position is randomly distributed in $[1,s_0]$, (2) each nonzero entry is uniformly distributed, i.e. $\{+1,-1\}$.


Defining $\textbf{A}=\textbf{S}_X^T\textbf{C} \in \mathbb{C}^{s_1 \times s}$ and $\textbf{B}=\textbf{S}_X^T\textbf{R} \in \mathbb{C}^{s_1 \times M}$, we finally obtain the factorized projection matrix via:
\begin{align}
\hat{\textbf{X}} =\argmin _{\text{rank}(\textbf{X})\leqslant K}||\textbf{A} \textbf{X} - \textbf{B}||_F^2= \pmb{\Omega}_A^{\dag} (\textbf{Q}_A^T\textbf{B}),
\end{align}
where $\texttt{qr}\{\textbf{A}\}=\textbf{Q}_A \pmb{\Omega}_A$ denotes the QR decomposition on $\textbf{A}$.
Thus, according to the approximated LRA, i.e. $\textbf{R}\simeq \textbf{C} \hat{\textbf{X}}$, we are able to compute the approximated $K$-rank SVD via:
\begin{align}
\hat{\textbf{R}}_K &\simeq  \textbf{C} \hat{\textbf{X}} = \textbf{C} \pmb{\Omega}_A^{\dag} (\textbf{Q}_A^T\textbf{B})_K, \nonumber \\
& = \textbf{C} \pmb{\Omega}_A^{\dag} (\textbf{U}_B \pmb{\Sigma}_B \textbf{V}_B^H) = \bar{\textbf{U}}  \bar{\bm{\Sigma}} \hat{\textbf{V}}^H \textbf{V}_B^H=  \bar{\textbf{U}}  \bar{\bm{\Sigma}} \bar{\textbf{V}}^H.
\end{align}
In this way, we acquire an approximated estimation of $\textbf{R}_K$, i.e. $\hat{\textbf{R}}_K$, via the above eq. (14).

When the joint Gaussian projection and count sketching are used to compute a factorized matrix $\hat{\textbf{X}}$, we will have the following theoretical result.

\begin{theorem}
Let $\textbf{C}=\textbf{RS}$ be the low-dimensional sketch of a large SPSD matrix $\textbf{R}$, with $\textbf{S}$ the Gaussian projection. $\textbf{Q}_C$ is the orthonormal bases of $\textbf{C}$, and $\hat{\textbf{X}}$ is the factorized projection in Eq. (13), provided a Gaussian+count sketching matrix $\textbf{S}_X\in \mathbb{R}^{M \times s_1}$. If $s_0=\mathcal{O}(K/\varepsilon+K^2)$ and $s_1=\mathcal{O}(K/\varepsilon)$ ($\varepsilon$ denotes the relative error, which is a vary small term), then we have
\begin{align}
||\textbf{Q}_{\textbf{C}} \hat{\textbf{X}} - \textbf{R}||_F^2 \leq (1+\varepsilon)^3||\textbf{R}-\textbf{R}_K||_F^2.
\end{align}
\end{theorem}

The above theorem can be easily proved by the {\it subspace embedding} property of Gaussian projection + count sketch \cite{clarkson2013low}. As seen, we can obtain the approximated SVD via randomized LRA, which leads to the $(1+\varepsilon')$ relative-error (with regards to the optimal $K$-rank SVD).

\begin{algorithm}[t]
\caption{Fast R-MUSIC}
\label{alg::conjugateGradient}
\begin{algorithmic}[1]
\Require
$\textbf{R}$: signal covariance matrix;
$K$: rankness of covariance matrix;
($s$,~$s_0$,~$s_1$): matrix sketching lengthes;
$\Delta \theta$: step size of angular griding.
\Ensure
Approximated signal-subspace $\hat{\textbf{U}}_K$;
Estimated DoAs of multiple targets $\pmb{\hat{\theta}}=\{\hat{\theta}_k\}~(k=1,2,\cdots,K)$
\State initial a sketching matrix $\textbf{S}\in \mathbb{R}^{M \times s}$ and derive a sketched matrix $\textbf{C} = \textbf{R} \textbf{S} \in \mathbb{C}^{M \times s}$;
\State initial another sketching matrix $\textbf{S}_X = \textbf{S}_C \times \textbf{S}_G \in \mathbb{R}^{M \times s_1}$;
\State compute two sketched matrices $\textbf{A}=\textbf{S}_X^T\textbf{C} \in \mathbb{C}^{s_1 \times s}$ and $\textbf{B}=\textbf{S}_X^T\textbf{R} \in \mathbb{C}^{s_1 \times M}$;
\State compute the QR decomposition $\texttt{qr}\{\textbf{A}\}=\textbf{Q}_A \pmb{\Omega}_A$; and then compute the projection matrix $\hat{\textbf{X}}  = \pmb{\Omega}_A^{\dag}\textbf{Q}_A^T\textbf{B}$;
\State compute the approximated SVD via: $\hat{\textbf{R}}_K  \simeq  \textbf{C} \hat{\textbf{X}}  =  \bar{\textbf{U}}  \bar{\bm{\Sigma}} \bar{\textbf{V}}^H$;
\State assign the signal subspace $\hat{\textbf{U}}_K \simeq \bar{\textbf{U}}(:,1:s)$; and for a given angle grid $\{0,\Delta \theta ,2\Delta \theta,\cdots,\pi\}$ compute the spatial pseudo-spectrum $\hat{P}_{\text{LRA}}(\theta) = \frac{1}{\textbf{a}^H(\theta)\left[\textbf{I}_M-\tilde{\textbf{U}}_K \hat{\textbf{U}}_K^H\right] \textbf{a}(\theta)}$;
\State estimate $K$ DoAs of unknown targets by identifying the peaks of $\hat{P}_{\text{LRA}}(\theta)$, i.e. $\hat{\theta}_{\text{LRA}} = {\texttt{peak}}_{\theta_k \in [0,\pi]}\{\hat{P}_{\text{LRA}} (\theta)\}.$
\end{algorithmic}
\end{algorithm}

\subsection{Fast Randomized MUSIC}

With the help of randomized LRA and approximated SVD of $\textbf{R}$, we now estimate the signal subspace by $\hat{\textbf{U}}_K \simeq \bar{\textbf{U}}(:,1:s)$;
in practice, we may configure $s\geq\hat{K}>K$.
On this basis, we are allowed to approximate the spatial pseudo-spectrum via:
\begin{align} \label{eq:approx_spectrum}
\hat{P}_{\text{LRA}}(\theta) = \frac{1}{\textbf{a}^H(\theta)\left[\textbf{I}_M-\hat{\textbf{U}}_K \hat{\textbf{U}}_K^H\right] \textbf{a}(\theta)},
\end{align}
where $\textbf{I}_M$ denotes one $M\times M$ identity matrix.
The unknown DoAs of multiple targets are then extracted by $\hat{\theta}_{\text{LRA}} = {\texttt{peak}}_{\theta_k \in [0,\pi]}\{\hat{P}_{\text{LRA}} (\theta)\}.$
Note that, repeating eq. \eqref{eq:approx_spectrum} for every $\theta \in [0, \pi]$ is also expensive \cite{Li2020Pseudospectrum}.
Fortunately, this can be computed in parallel by using multiple processors.
The fast evaluation of eq. (\ref{eq:approx_spectrum}) will be left to the future study.

\subsection{Time Complexity of Subspace Computation}

Based in the above analysis, a schematic flow of our R-MUSIC method is given in \textbf{Algorithm 1}.
Obviously, the complexity of our R-MUSIC method involves three parts: (i) determining a low-dimensional abstract $\textbf{C}$, (ii) solving eq. (13), (iii) computing the approximated SVD via eq. (14).
In particular, once two small factorized matrices $\textbf{C}$ and $\textbf{R}$ are attained, the time complexity in approximating SVD of $\textbf{R}$ is reduced to $\mathcal{O}\{s^2 M\}$ as Eq. (14); whilst an exact computation requires a complexity $\mathcal{O}\{M^3\}$.
Accordingly, the whole complexity of R-MUSIC is $\mathcal{O}\{\text{nnz}(\textbf{R}) + M \cdot \text{poly}(K/\epsilon)\}$, whereby $\text{nnz}(\textbf{R}) \leq M^2$ is the number of non-zero entries in $\textbf{R}$; and $\text{poly}(K/\epsilon)$ represents a polynomial of $K/\epsilon$.
So, the time complexity in computing an approximated subspace is dramatically reduced, when compared to previous full-SVD and exact $K$-SVD of $\textbf{R}$.


\section{Performance Simulation \& Analysis}
In the section, we provide the numerical simulation and performance analysis of our R-MUSIC approach in the context of massive MIMO radars.
As found, since there is no loop/recursion in our scheme, the computational complexity can be alternatively evaluated via the mean run-time (basic CPU frequency 3.5GHz, 4GB memory), which is rigorously proportional to the number of multiplication flops.

\subsection{Time Complexity}
We firstly evaluate the time complexity of various methods in acquiring the signal/noise sub-space, which constitutes a critical step to super-resolution radar signal processing.
In the analysis, we fix $N=M$ and $K=9$.
The signal-to-noise-ratio (SNR) is defined as $\mathbb{E}\{s^2(t)\}/\sigma_n^2$ and is configured to be 5dB.
When deriving a low-dimensional sketch $\textbf{C}$, we assume $s=K$.
Note that, in order to achieve the upper bound on the LRA error as in Eq. (15), two parameters in a compound sketching matrix $\textbf{S}_X$ should meet: $s_0=\mathcal{O}(K/\varepsilon+K^2)$ and $s_1=\mathcal{O}(K/\varepsilon)$.
From the numerical analysis, unlike the strict theoretical result, we may directly set $s_0=2K$ and $s_1=(1+\eta)\times K~(0<\eta<1)$.
As demonstrated, this would suffice to produce the accurate DoA estimation in practice.


\begin{figure}[!t]\vspace{0pt}
	\begin{center}
		\subfigure[ ] {
			\includegraphics[width=70mm]{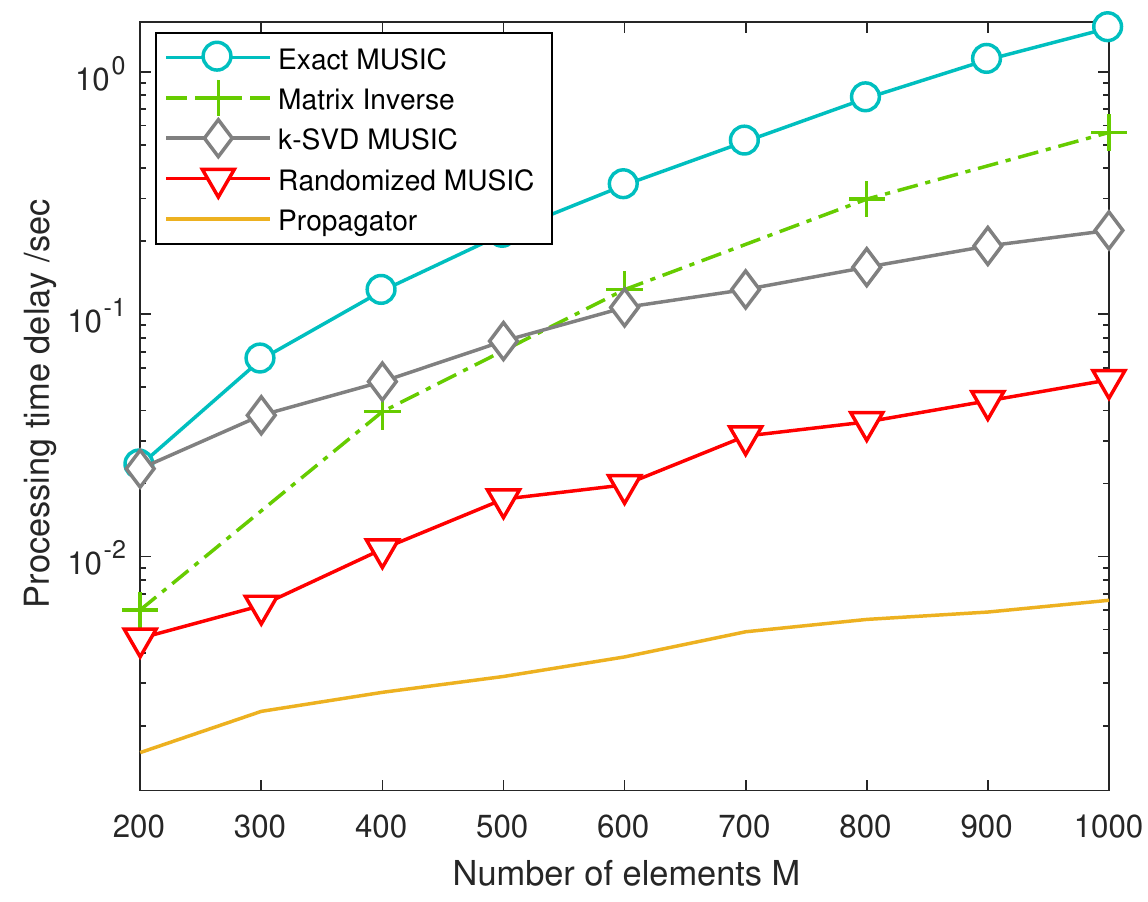}}
		\subfigure[ ] {
			\includegraphics[width=70mm]{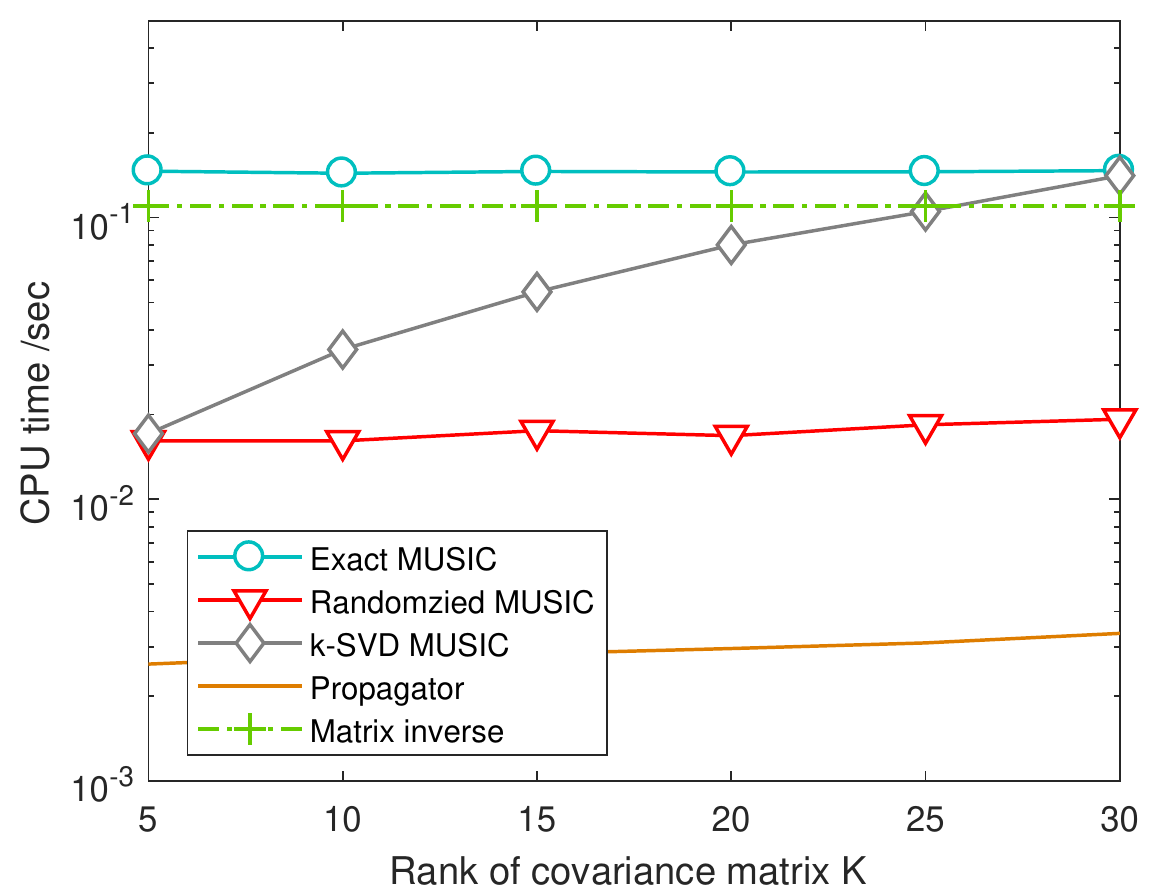}}
		\caption{Time complexity of various methods. (a) Processing delay of different $M$ ($K=9$). (b) Processing delay of different $K$ ($M=700$).
        }
	\end{center}
\end{figure}

\begin{figure}[!t]
\centering
\includegraphics[width=7cm]{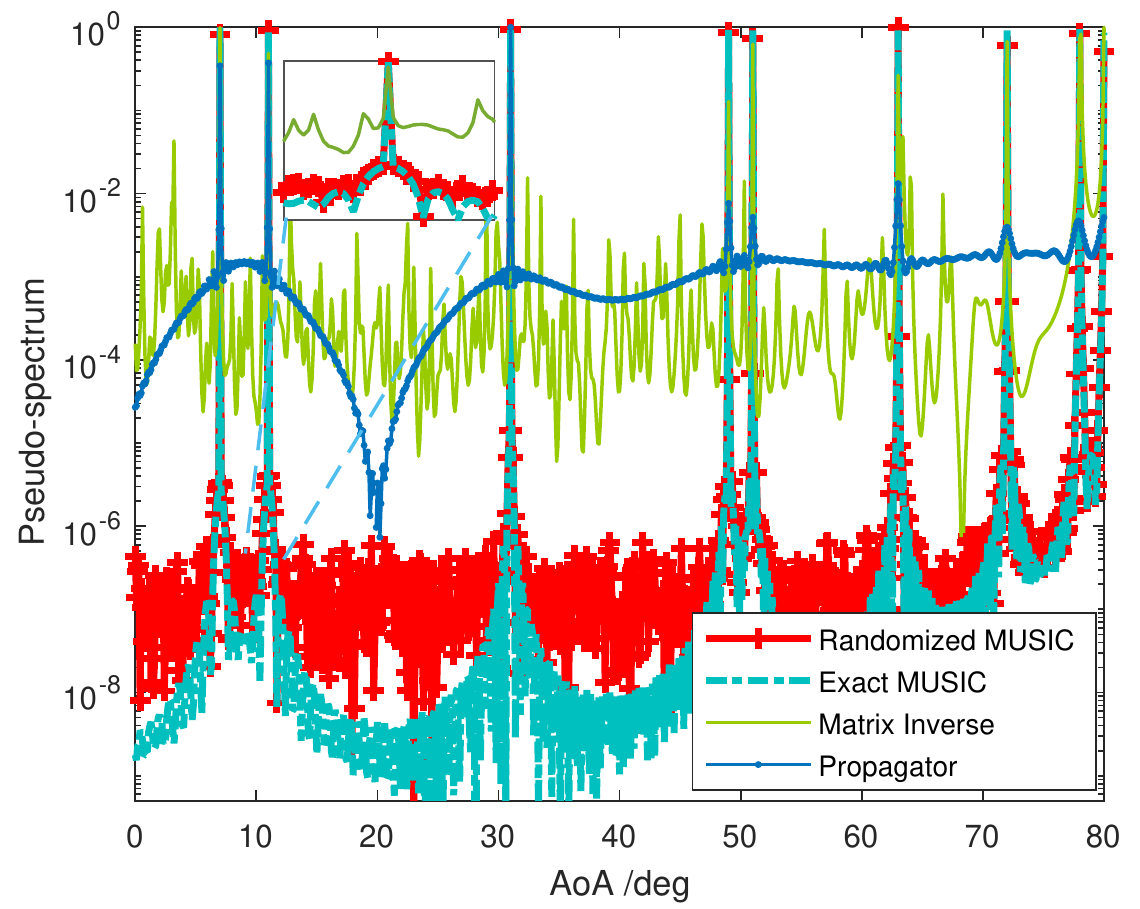}
\caption{Estimated pseudo-spectrum of different methods ($M=300$, $K$=9 and SNR=$-5$dB).
}
\end{figure}

From Fig. 2-(a), the full SVD, in the context of massive MIMO radars, tends to be practically infeasible, due to its unaffordable computation, which would also incur a noticeable processing delay.
When $M=1000$, its processing time will be 1.52 sec, which miserably fails to meet the requirement of real-time sensing.
For the other exact $K$-SVD method \cite{Grimes1994A}, the time complexity is $\mathcal{O}\{KM^2\}$, which also tends to be inapplicable to massive receiving elements ($M>$200), e.g. producing also a remarkable delay (i.e. $0.22$ sec).
As far as its lowest complexity is concerned, the Propagator method is more promising to massive MIMO radars.
For example, its complexity is linearly scaled as $M$ grows, which thus enables the real-time signal processing in automotive scenarios.

We also evaluate the time complexity of various methods for a fixed massive MIMO radar, e.g. $M=700$.
In the simulation, we assume the rank of a covariance matrix (i.e. $K$) will be changed from 5 to 30.
As observed, in this situation the time complexity of exact MUSIC, Propagator and our R-MUSIC method will be only slightly increased.
In contrast, the complexity of the other popular $K$-SVD method grows rapidly.
In this regard, the computational advantage of our R-MUSIC method shall become more obvious.

By utilizing a new paradigm of randomized sketching and approximated computing (rather than the existing deterministic computing), our R-MUSIC approach would effectively save the computational resource and reduce the time delay in massive MIMO radars.
To be specific, now the processing delay of R-MUSIC is only 0.04 sec, which is remarkably faster than full-SVD or exact $K$-SVD methods.
Consequently, it potentially permits the real-time environment sensing and target estimation in various automotive applications \cite{Li2019Fast}.

\subsection{DoA Estimation Accuracy}
In Fig. 3, we further examine the resolution of pseudo-spectrum approximated by our R-MUSIC algorithm.
In this numerical analysis, we assume $N=M=300$ and SNR=$-$5dB.
We observed that our approximated pseudo-spectrum in eq. (16) is as accurate as that of the near-optimal MUSIC scheme derived by the full-SVD in eq. (6).
The only difference in our R-MUSIC scheme is that, as highlighted in sub-figure, the side-lobe and the noise floor would be slightly increased in its approximated pseudo-spectrum.
Even so, unknown DoAs of multiple targets should be perfectly identified.
For another Propagator method, however the estimation error would be inevitably, especially when unknown targets fall into special angular region (i.e. [70, 90] and [-90, -70]), no matter what the estimation SNR is.
As such, our R-MUSIC method is extremely attractive to the high-resolution yet real-time environment sensing (e.g. target detection/estimation) in massive MIMO radars.
\begin{figure}[!t]
\centering
\includegraphics[width=7cm]{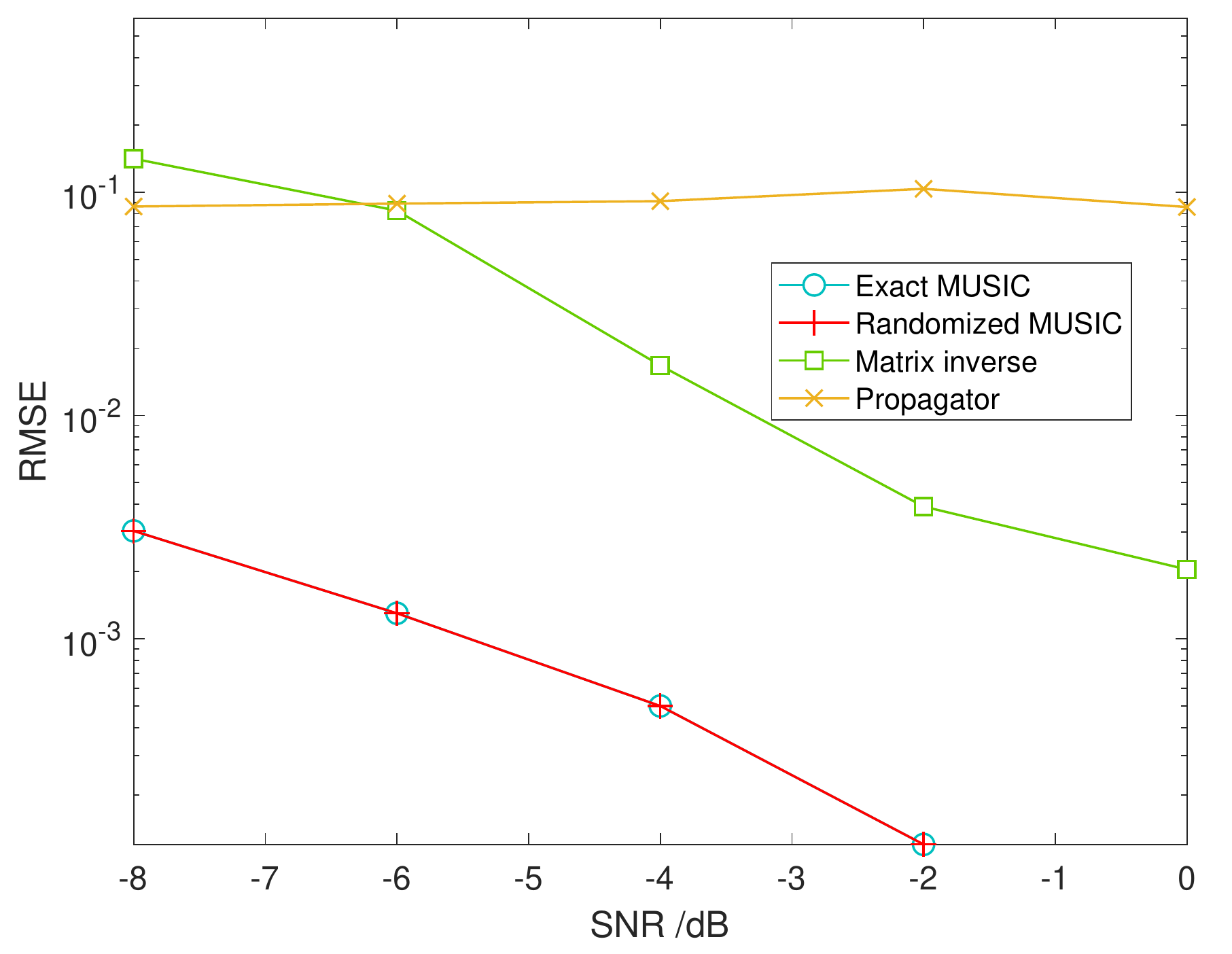}
\caption{Root-mean-square error of various algorithms ($M=200$, $K$=9).}
\end{figure}

Finally, we investigate the root mean square error (RMSE) performance of various DoA estimation methods.
From Fig. 4, both the matrix inverse method and the Propagator method attain the less accurate DoAs estimations.
In particular, since the estimation error in Propagator persists when unknown DoAs fall into [70, 90] deg (or [-90,-70] deg) \cite{tayem2005l-shape} (see also Fig. 3), the RMSE floor seems to be inevitable.
Second, our R-MUSIC approach acquires the comparable accuracy as the exact MUSIC method.
This is relatively easy to follow. Recall in Eq. (15), when the rank of a covariance matrix $\textbf{R}$ is $K$, the randomized decomposition $\textbf{C}\textbf{X}$ would accurately approximate $\textbf{R}$ (i.e., the residual term $||\textbf{R}-\textbf{R}_K||_F^2$ may approach zero).
As a result, the estimated signal sub-space should accurately approximate that of the MUSIC method, and therefore the near-optimal DoA estimation can be attained (see the approximated pseudo-spectrum in Fig. 3).
To sum up, our R-MUSIC method coyld produce the super-resolution estimation as MUSIC, whilst significantly reducing the time complexity of massive MIMO radars.

\section{Conclusion}
Relying on a novel concept of randomized matrix sketching, the computational efficient estimation of signal subspace is studied for high resolution massive MIMO radars.
By exploiting the inherent low-rank property of a covariance matrix in automotive applications, we obtain one approximated signal subspace via our R-MUSIC method.
Relying on the approximated (rather than exact) matrix computation, our R-MUSIC accomplishes low-complexity yet high-performance DoA estimations, which provides one promising solution to automotive radars.
As shown, our fast R-MUSIC offers the significant advantage over other popular candidates, e.g. reducing the time-complexity by tens of times.
More importantly, it attains the same accuracy as the near-optimal MUSIC method.
Thus, our R-MUSIC method resolves the inherent contradiction between complexity and accuracy, which paves a potential way for the high-resolution real-time radar signal processing in emerging automotive applications.

\section*{Acknowledgment}
This work of B. Li and C. L. Zhao was supported by Natural Science Foundation of China (NSFC) under Grants U1805262.

\bibliography{scibib}

\bibliographystyle{ieeetr}

\end{document}